\documentclass{WileyMSP-template}

\usepackage {mathtools}
\usepackage{bm}
\usepackage{color}
\usepackage{pdfpages}

\begin{document}

\pagestyle{fancy}
\rhead{\includegraphics[width=2.5cm]{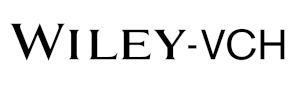}}

\title{Pulse dynamics of flexural waves in transformed plates}

\maketitle

% Author: Please give full first and last names for authors and include * after the name of all corresponding authors

\author{Kun Tang}
\author{Chenni Xu}
\author{S\'ebastien Guenneau}
\author{Patrick Sebbah*}

% Dedication

%\dedication{Optional dedication here. If no dedication is required, please leave blank}

% Affiliations: Please provide adacemic titles (Prof. or Dr.) for all authors where applicable, and include an institutional email address for all corresponding authors
\begin{affiliations}

Dr. Kun Tang, Dr. Chenni Xu, Prof. Patrick Sebbah\\
Department of Physics, The Jack and Pearl Resnick Institute for Advanced Technology, Bar-Ilan University, Ramat-Gan 5290002, Israel.\\
Email Address:patrick.sebbah@biu.ac.il

Dr. Chenni Xu\\
%Department of Physics, The Jack and Pearl Resnick Institute for Advanced Technology, Bar-Ilan University, Ramat-Gan 5290002, Israel.\\
Zhejiang Province Key Laboratory of Quantum Technology and Device, Department of Physics, Zhejiang University, Hangzhou, 310027, Zhejiang, China.\\

Prof. S\'ebastien Guenneau\\
UMI 2004 Abraham de Moivre-CNRS, Imperial College London, London SW7 2AZ, United Kingdom.\\

%Prof. Patrick Sebbah\\
%Department of Physics, The Jack and Pearl Resnick Institute for Advanced Technology, Bar-Ilan University, Ramat-Gan 5290002, Israel.\\

\end{affiliations}

% Keywords: Please provide a minimum of three and a maximum of seven keywords, separated by commas


% Abstract should be written in the present tense and impersonal style (i.e., avoid we), and be at most 200 words long
\begin{abstract}
Coordinate-transformation-inspired optical devices have been mostly examined in the continuous-wave regime: the performance of an invisibility cloak, which has been demonstrated for monochromatic excitation, %would inevitably
is likely to deteriorate for short pulses. Here we investigate pulse dynamics of flexural waves propagating in transformed plates. We propose a practical realization of a waveshifter and a rotator for flexural waves based on the coordinate transformation method.
Time-resolved measurements reveal how the waveshifter deviates a short pulse from its initial trajectory, with no reflection at the bend and no spatial and temporal distortion of the pulse.
Extending our strategy to cylindrical coordinates, we design a wave rotator. We demonstrate experimentally how a pulsed plane wave is twisted inside the rotator, while its wavefront is recovered behind the rotator and the pulse shape is preserved, with no extra time delay. We propose the realization of the dynamical mirage effect, where an obstacle appears
oriented in a deceptive direction.
\end{abstract}

\section{Introduction}
In two independent proposals, Pendry et al. \cite{Pendry1780} and Leonhardt \cite{Leonhardt1777} showed that a transformation of coordinates can map a particular distortion of the electromagnetic field onto a change of material properties -inhomogeneous and anisotropic-, unveiling unlimited possibilities for the design of metamaterials with new functionalities to control the flow of light. This new concept was originally proposed to design an invisibility cloak that was first validated for electromagnetic waves \cite{Schurig977} and thereafter extended to other types of waves including acoustic \cite{Cummer_2007}, hydrodynamics \cite{hydrodynamic_cloak} and water waves \cite{water_wave_cloak}. In all these cases, the governing equations are form invariant. When one moves to the area of elastic waves however, the elasticity equations are in general not form-invariant under a general coordinate transformation \cite{Milton_2006}, %\cite{Norris}
except in the framework of Cauchy elasticity \cite{Nassar_PRSA,Nassar_PRL,brun_apl} and in the framework of Willis materials \cite{Norris_2011,Achaoui2020}. Consequently, if cloaking exists for such a class of waves, it would be of a different nature than its acoustic \cite{Norris} and electromagnetic counterparts \cite{Nassar_PRL}. Researchers resort to studying the special case of flexural waves in thin plates, which are described by the the fourth order Kirchhoff-Love equation. Over the past ten years, there have been various theoretical proposals for the design of elastic invisibility cloaks for flexural wave \cite{PRB_farhat,PRL_farhat,COLQUITT2014131,Brun_2014,PRE_darabi,arxiv_pomot}, followed by their experimental validations \cite{cloaking_wegner,cloaking_darabi,Misseroni2016}. %illusion_jensen,Buckmann4930,Buckmann2014
However, experimental investigations of cloaking have been mostly restricted to continuous-wave (CW) excitation \cite{Schurig977,optical_cloaking,cloaking_wegner,cloaking_darabi,2D_ground_cloaking,Cloaking_nico_fang}. Cloak invisibility to short pulses has been rarely tested \cite{pulse_LC,2D_ground_cloaking,PNAS_surface_wave,SR_surface}, since these inhomogeneous, magnetic, and anisotropic metamaterial structures are subject to inherent frequency dispersion, which is likely to distort the pulse both in space and time and makes its reconstruction challenging \cite{optic_delay,temporal_boris}. Broadband cloaking has been realized in various systems, including acoustic \cite{Cloaking_nico_fang}, elastic \cite{cloaking_wegner,cloaking_darabi}, and water waves \cite{water_wave_cloak}. But achieving broad spectral range operation does not guarantee that a pulse propagating through the transformed medium will remain undistorted.

The coordinate transformation used in the design of electromagnetic cloaks with cylindrical geometry leads to a gradient distribution of permittivity and permeability, which necessitates engineering the magnetic response of materials hardly available in the optical range \cite{Pendry1780,Schurig977}. To solve this issue, a reduced set of parameters has been proposed, which provides with non-magnetic structures, while preserving the cloaking performances, but which suffers from reflection and scattering due to impedance mismatch at the outer boundary \cite{optical_cloaking}. Another approach to avoid magnetic materials and preserve the invisibility of the device itself is to maintain the determinant of the Jacobian transformation tensor at unity, therefore preserving the volumes throughout the space \cite{Non_magnetism,han_ol,xu_josa,han_oe}. Compared to cylindrical cloaking, this so-called non-magnetic geometrical transformation is continuous i.e. adiabatic \cite{shifter_huanyang,cai_apl}. The space is not abruptly stretched or compressed and the topology is conserved during the transformation process \cite{shifter_rahm}, ensuring perfect impedance matching at the boundaries. This volume-preserving method has never been considered for elastic waves, where it could meet the challenge of designing intrinsically reflection-less elastic devices \cite{cloaking_wegner,cloaking_darabi,Misseroni2016}. We note that our adiabatic route to elastodynamic cloaking is markedly different from the design of elastostatic cloak via a direct lattice transformation \cite{Buckmann4930}, which was recently extended to the dynamic regime \cite{kadic}.

Transformation optics actually has not been limited to the design of invisibility cloaks but has led to the development of novel wave-manipulation devices \cite{concentrotor_huanyang,rotator_huanyang, wormhole_jensen,shifter_rahm}. Among them, the waveshifter, the building-block of fundamental steering optical components such as the wave splitter, and the rotator, a %invisible
device capable of twisting and restoring waves, creating at the same time a mirage effect \cite{zolla_mirage}, have been proposed to control electromagnetic waves \cite{rotator_huanyang}, as well as scalar acoustic waves \cite{shifter_jensen,rotator_jiang}, water waves \cite{shifter_water,rotator_water} or hydrodynamic flows \cite{rotator_park}. Surprisingly, these new classes of devices have never been considered for elastic waves.

In this article, we adapt the non-magnetic geometrical transformation to elastic waves and call on for its continuous and volume preserving character to design a reflection-less waveshifter and a %invisible
wave rotator for flexural waves. Here, we investigate the pulse dynamics of flexural waves propagating inside a 3D-printed transformed plate by mapping the spatio-temporal field distribution in response to a short pulse excitation. The elastic pulse is shown to perfectly follow the $20^{\circ}$ bent of the waveshifter with minor back reflections at the bend and negligible spatial and temporal dispersion of the incident short pulse. A modal analysis shows how higher modes of the bent waveguide are prevented from being excited.
%\textcolor{red}{We illustrate the importance of such a transformed device by using it as the building block of a truly invisibility cloak for elastic pulses.}
The elastic wave rotator is designed following the same strategy. Most remarkably, a short pulse plane-wave excitation is shown to be restored, after experiencing a $30^{\circ}$ twist within the rotator. The pulse is shown to cross the device with no extra time delay, as if it was propagating through ``free space''. To the best of our knowledge, this has never been observed before in the context of elastic waves, and never in the time domain with electromagnetic waves \cite{rotator_huanyang}, acoustic waves \cite{shifter_jensen,rotator_jiang}, water waves \cite{shifter_water,rotator_water} or hydrodynamic flow \cite{rotator_park}.
Three dimensional full-elasticity simulations support our experimental observations. Finally, this rotator is proposed to demonstrate the mirage effect with elastic waves, where a scattering object is seen to radiate from a deceptive direction.

\section{Waveshifter}
\begin{figure*}
\centering
\includegraphics[width = 16cm]{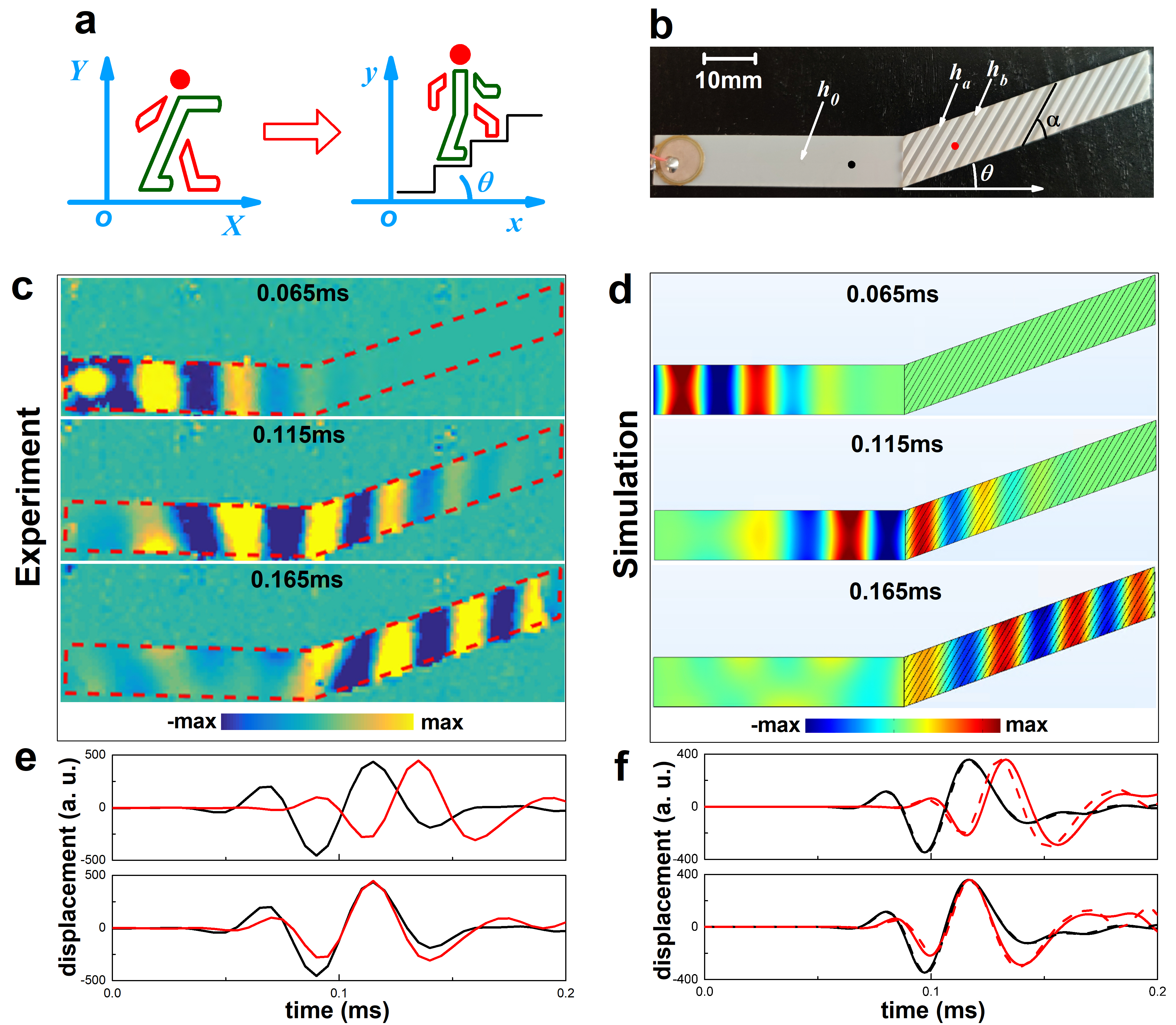}
\caption{\textbf{Dynamics of pulse propagation in the waveshifter.} (a) Sketch of the coordinate mapping from a horizontal waveguide in virtual space (X,Y) to an oblique waveguide in the real space (x, y). (b) Top view of the waveshifter: the waveguide is a 100~mm-long, 10~mm-wide 3D-printed ceramic plate, with a bending angle $\theta=20^{\circ}$. Thickness is $h_0$= 0.5~mm for the left arm. The corrugation in the right arm (period 2~mm) alternates thicknesses $h_a$= 0.785~mm  and $h_b$= 0.318mm, and form an angle $\alpha=40^\circ$ with the waveguide axis. A piezoelectric diaphragm is attached to the left end of the waveguide. (c) Experiment: Snapshots of the out-of-plane velocity field in response to a Ricker pulse (see [Experimental Section]) with central frequency 20~kHz, measured at times 0.065~ms (top), 0.115~ms (middle), and 0.165~ms (bottom). The red dashed line outlines the physical limit of the structure. (d) Full-3D numerical simulations: Snapshots of the out-of-plane velocity field calculated at same times as (c), using Finite Element Method in Comsol Multiphysics. (e) Experiment: Pulse profile measured at two spatial positions, before and after the bend, marked in (b) by black and red dots, respectively. (Top) Actual measurement showing the time delay accumulated during propagation. (Bottom) The transmitted pulse (red) has been time-shifted to show the coincidence with the incident pulse (black). (f) Numerical simulations: Pulse profiles calculated at same positions as in (e). (Top) and (Bottom) same as in (e). The dashed red line is the pulse profile after propagation in a straight plain waveguide (no bend and no corrugation). The perfect overlap between the 2 red curves demonstrates that the residual deformation of the pulse is solely due to the natural dispersion of flexural waves in thin plates.}
\end{figure*}

In the limit of plate thickness much smaller than the wavelength, the phase velocity $c$ of flexural waves can be described within the Kirchhoff-Love plate theory as $c=\sqrt{\frac{D \omega}{\rho h}}$, where $D=\frac{E  h^3}{12 (1-\nu^2)}$ is the flexural rigidity of the plate, $\rho$ its mass density, $h$ its thickness and $E$ its Young's modulus, $\omega$ being the angular frequency. By analogy with layered anisotropic electromagnetic media \cite{shifter_huanyang}, anisotropic phase velocity for flexural waves can be achieved by alternating layers of materials with different elastic properties. A possible approach is to vary the Young's modulus or the density of the successive materials, as successfully demonstrated by the group of Wegener \cite{cloaking_wegner}. But from a practical point of view, varying the thickness of the plate, rather than modifying its intrinsic elastic parameters, turns out to be a much simpler strategy to introduce anisotropy. This approach has been validated for adiabatically varying thickness, e.g. for the design of elastic lenses \cite{fisheye,lenses_climente,Grin_lense}. Here, we demonstrate that, in contrast to an adiabatically varying thickness, a periodic change of plate thickness on a subwavelength scale can be utilized to design effective anisotropic metamaterials for flexural waves. A significant advantage in terms of realization is that a single material is required and that the subwavelength structuration is easily implemented by surface machining or, more conveniently,  by 3D printing, a technique now readily available for many types of materials, including ceramics and metals. Here we illustrate our novel approach with the design of a waveshifter.

Consider a walker progressing along a horizontal path. Further consider the space transformation which transforms this flat path into a staircase. The walker now finds himself going upstairs, at an angle $\theta$ with the horizontal axis, while maintaining an upright position, as if nothing had changed for him (see illustration in \textbf{Figure. 1(a)}). To describe the associated coordinate transformation, we consider the mapping which transforms a point $(X,Y)$ in the virtual space (the flat path) into a point $(x,y)$ of the oblique region in the real space (the staircase). This change of coordinates can be expressed as:
\begin{equation}
     \begin{cases}
x = X &  \\
y = X \tan \theta + Y
       \end{cases},
 \end{equation}
where $\theta$ is the steering angle.\\
The Jacobian Matrix of the above geometrical transformation is
\begin{equation}
  F =\left(
 {\begin{array}{ccccc}
   1 & 0\\
   t & 1\\
  \end{array}}
       \right),
%\label{eq:waves hifter_jaco}
\end{equation}
where $t= \tan \theta$. For a given angle $\theta$, $F$ is independent of space coordinates. Note also that the determinant is unity, $J=\mid detF \mid=1$, so that the transformation we have defined preserves the area throughout space.

Based on geometrical transformation method, mapping of coordinates from the homogeneous virtual space to the real space, results in a change of material parameters in the transformed wave equations. In the tracks of  \cite{PRB_farhat,arxiv_pomot}, the transformed Kirchhoff-Love equation results in an anisotropic flexural rigidity of the form
\begin{equation}
  \overline{D}
       = D_0 F F^T F F^T J^{-2}.
%\label{eq:waveshifter_trans}
\end{equation}
The flexural rigidity becomes tensorial and accounts now for the anisotropy of the transformed medium. In contrast to elastic cloak where it varies radially \cite{cloaking_wegner,cloaking_darabi}, the rigidity tensor in our case is independent of the space coordinates. The mass density itself, $\overline{\rho}=\rho_0J^{-2}=\rho_0$, remains unaffected by this coordinate transformation. This means that the Kirchhoff-Love equation in time domain is form-invariant, where $\rho$ is a factor of the second time derivative of the Kirchhoff-Love equation. This greatly simplifies its implementation in an actual device, as we proceed to explain.

\begin{figure*}
\includegraphics[width = 16cm]{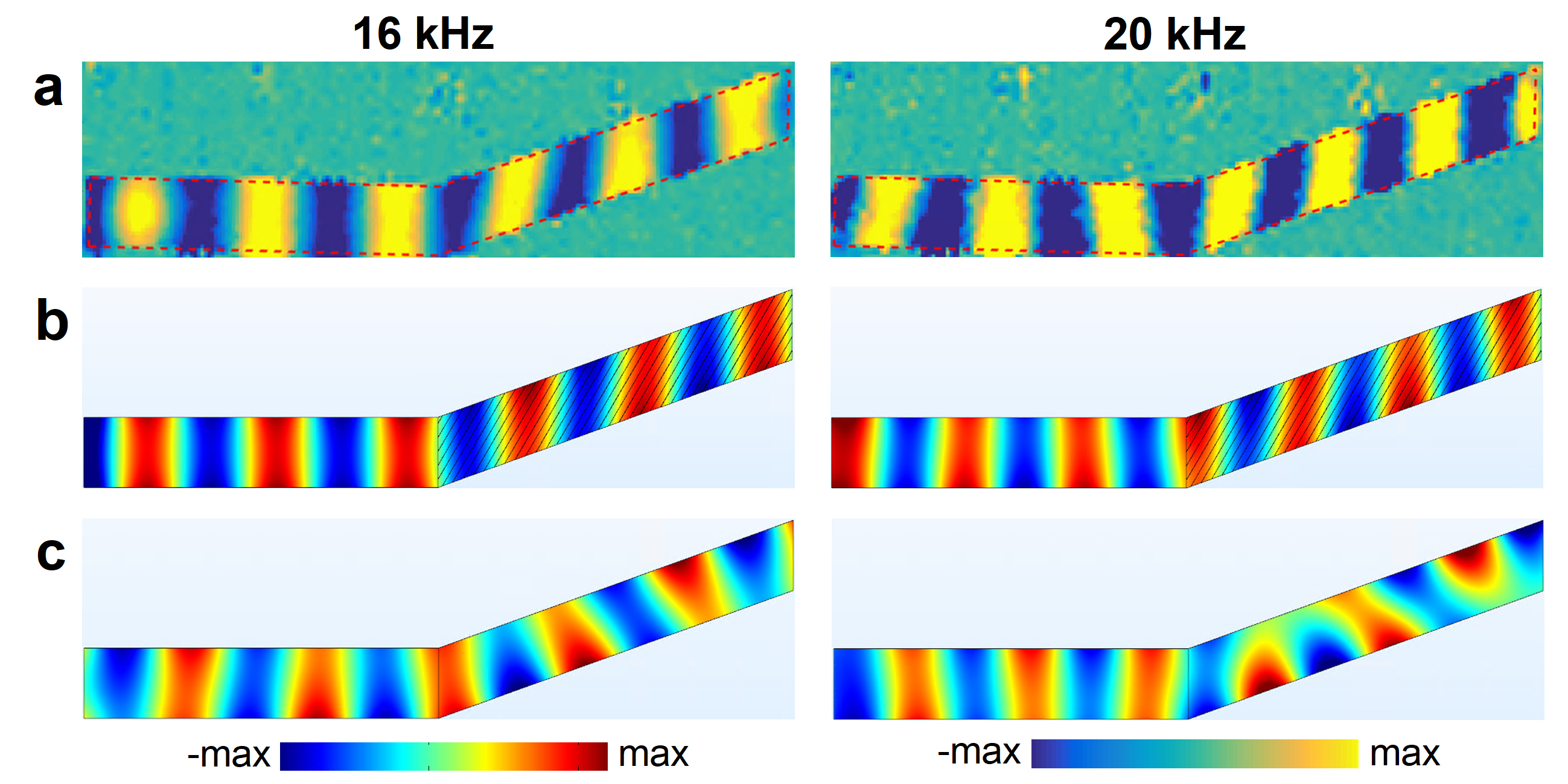}
\caption{\textbf{Single-frequency wave steering across the waveshifter.} (a) Experimentally measured out-of-plane velocity field at two different frequencies 16kHz (left column), and 20kHz (right column); (b) Corresponding  numerical results; (c) For comparison, the numerical simulations are shown for an empty bent waveguide (no metamaterial structure).}
%\label{fig:}
\end{figure*}

A medium with this particular anisotropic rigidity $\overline{D}$ can then be realized with a simple subwavelength structure by invoking effective medium theory \cite{milton_2002}. Here we propose to approximate the homogeneous anisotropic medium by a bi-layered structure consisting of an alternation of two materials with identical widths but different flexural rigidities $D_a$ and $D_b$. The normal to the layered structure defines the direction of anisotropy. If in addition this direction makes an angle $\alpha$ with the direction of propagation, the effective rigidity tensor is then given by
\begin{equation}
  \overline{D}_{oblique}
  = R_{\alpha}^T \left[
 {\begin{array}{ccccc}
   \ D_{\parallel} & 0\\
   0 & D_{\perp} \\
  \end{array}}
       \right] {R_{\alpha}},
      % \label{eq:waveshifter_obliq}
\end{equation}
with $D_{\parallel}=2D_a D_b/(D_a+D_b)$, $D_{\perp}=(D_a+D_b)/2$ and $ R_{\alpha} = \left[
 {\begin{array}{ccccc}
   \cos\alpha & -\sin\alpha\\
   \sin\alpha & \cos\alpha\\
  \end{array}}
       \right] $ the conventional rotation matrix.\\
Identifying Equations. (3) and (4), we obtain the rigidity profile for the transformed region, and a general expression for the angle $\alpha$, $D_a$ and $D_b$, in terms of parameter $t=\tan \theta$:
\begin{equation}
\left\{
\begin{array}{lr}
      t = 2\cot {(2 \alpha)} \\
      D_a / D_0 = \cot^4 (\alpha)+ \sqrt{-1+\cot^8{(\alpha)}}       \\
      D_b / D_0 = \cot^4 (\alpha)- \sqrt{-1+\cot^8{(\alpha)}}       \\
\end{array}.
\right.
\end{equation}

Keeping all elastic parameters constant, the rigidity for each layer, $D_a$ and $D_b$, is easily implemented by adjusting the  thickness, $h_a$ and $h_b$, of the plate at each layer, following the definition of $D$ \cite{cloaking_darabi,lenses_climente,fisheye,Grin_lense}.
According to Equation.~(5), the thickness difference $h_a-h_b$ increases steeply for $\theta$ above $40^{\circ}$ (see Figure. S1 in [Supplementary Information]). We therefore choose for the experimental realization an angle $\theta=20^{\circ}$, in order to limit the geometrical constraints for the 3D-printing. Nevertheless, the design is shown in the [Supplementary Information] to be effective up to $\theta=60^{\circ}$.
%We refer to Figure. S1 in [Supplementary Information] for required plate rigidity and thickness in each layer depending upon the deviation angle $\theta$. Opting for a variation of rigidity single material can therefore be used, making 3D-printing particularly well-suited to structure the plate and to design the transformed device, provided the angle $\theta$ is not too large: The contrast in layers' thicknesses increases steeply when $\theta$ is above $40^{\circ}$. We find that $\theta=20^{\circ}$ is a good trade-off between easiness in 3D printing and a marked cloaking effect.

Following our method, we design a waveguide with total length $L$=100~mm, width $W$=10~mm, and a deviation angle $\theta$=$20^{\circ}$, as shown in Figure. 1(b). In the left section of the waveguide, plate thickness is $h_0$=0.5~mm, while the bent (right) section is corrugated at an angle $\alpha$=$40^{\circ}$ with respect to the waveguide axis, as calculated from Equation.~(5) and shown in Figure. 1(b). The period of the square corrugation is 2~mm, with alternating plate thicknesses $h_a$=0.785~mm and $h_b$=0.318~mm, to be compared with the wavelength, $\lambda$=18.7~mm at 16~kHz. We chose deliberately to design our waveshifter in a waveguide geometry (rather than ``free space" propagation) in order to investigate the modal coupling at the bend while the energy flow is steered in the oblique direction, as this will be discussed later. Figure 1(b) depicts the waveshifter device, 3D-printed in Zirconium dioxyde ceramic. To investigate the pulse dynamics along the bent waveguide, we examine its temporal response to a Ricker pulse (see definition in [Experimental Section]) centered at 20~kHz, launched from its left section. A laser vibrometer is used to map the spatio-temporal evolution of the elastic field along the waveguide. The experimental setup is described in full details in [Experimental Section]. In Figure. 1(c), we present snapshots of the velocity field distribution at three different times. The pulse is seen to propagate smoothly without deformation along the waveguide while it is deflected, with almost no reflection at the waveguide bend. Most striking, the wavefront remains vertical, in the same direction as the incident wave, like the walker in Figure.~1(a). This is the main feature expected from the waveshifter: The anisotropic section of the waveguide as we designed it maintains perfectly the initial wavefront direction, giving the illusion that the wave comes from the same direction, although its energy has been deviated. This perfectly realizes the mapping of Figure. 1(a). The experiment is successfully compared against time-domain full-3D elastic wave simulations, as shown in Figure. 1(d).  The complete movie of the pulse propagation is available in [Supplementary Information].

The investigation of the pulse dynamics reveals however a new remarkable feature of the waveshifter: ``that the structuration of the plate does not introduce any additional spatial and temporal dispersion, in addition to the natural dispersion of flexural waves.'' %that spatial and temporal dispersion of the pulse is negligible during propagation and deflection.
To demonstrate it, we measure the pulse profile before and after the bend, at positions indicated by black and red dots in Figure.~1(b). Figure.~1(e) shows the pulse before (top) and after (bottom) time-shifting the transmitted pulse (red). We find that the two signals almost perfectly overlap. Surprisingly, the coincidence between the two signals turns out to be even better in the experiment than in the numerical simulations shown in Figure.~1(f). Actually, the small residual temporal deformation observed in Figure. 1(e) and 1(f) is solely due to the natural dispersion of the flexural waves during propagation. This is demonstrated in Figure. 1(f), where a comparison is made with a pulse propagating through an empty straight waveguide (red dashed line). If a time delay exists between ``free'' propagation and propagation in the waveshifter (top), the two pulses perfectly overlap (bottom), showing that there is no pulse distortion due to the transformed device.

\begin{figure*}
\centering
\includegraphics[width=16cm]{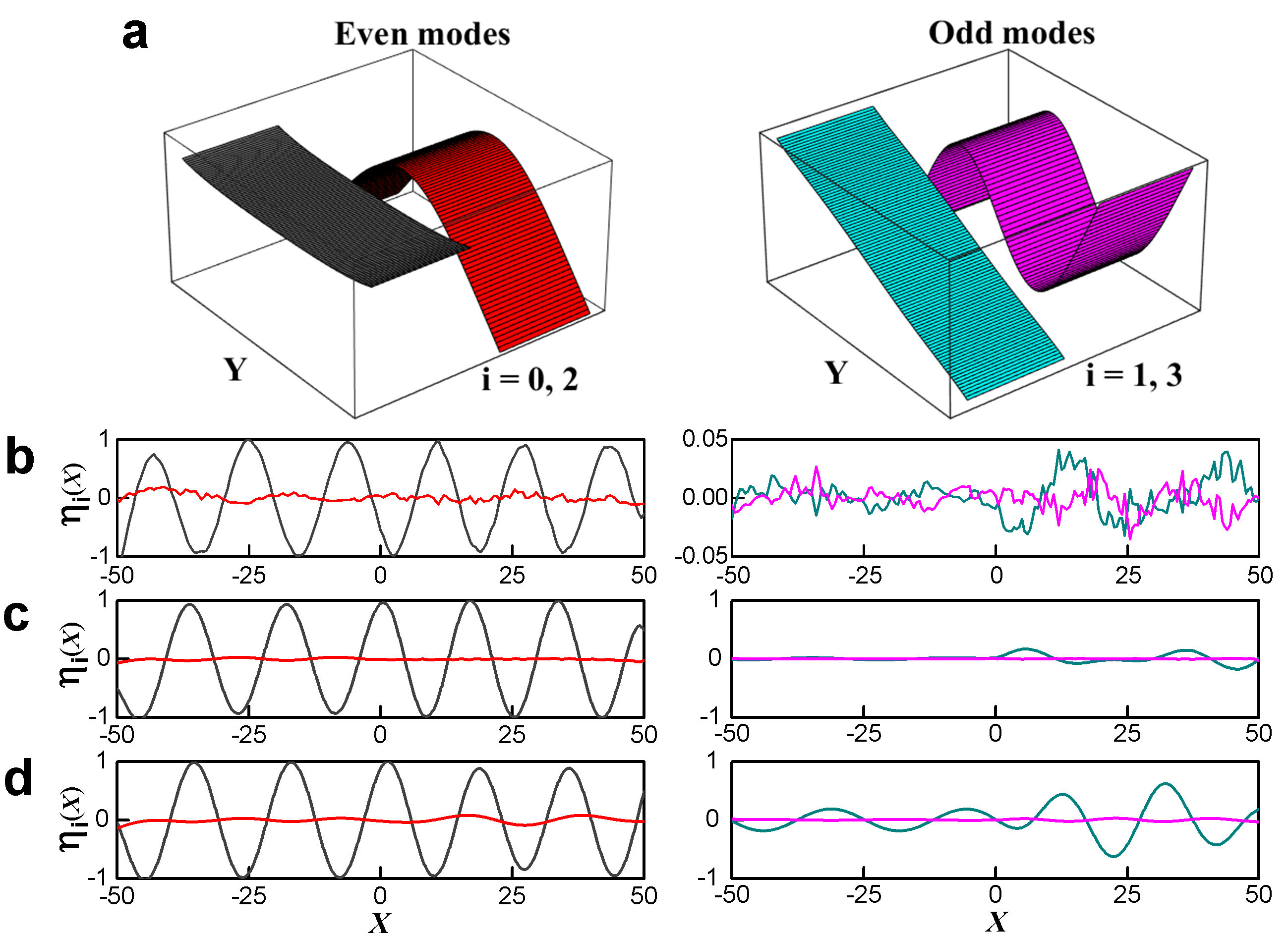} % was 8.5cm
\caption{\textbf{Modal analysis before and after the bend at frequency 16~kHz.} (a) Mode profiles (y-dependence) of the first two even (left panel) and odd (right panel) flexural eigenmodes. Black and red colors correspond to the $0^{th}$ and $2^{nd}$ even modes, while cyan and magenta correspond to the $1^{st}$ and $3^{rd}$ odd modes. Modal components $\eta_i (X)$ ($i=0,1,2,3$) along the waveguide, before and after the bend ($X=0$), (b) for measured results, (c) for numerical simulations and (d) for the empty bent waveguide without corrugations. The left panels describe even modes, while the right panels are for the odd modes. (Black) zeroth-order modes; (red) second-order modes; (cyan) first-order modes; (magenta) third-order modes; Each row is normalized with the maximum amplitude of the $0^{th}$-order even mode, $\eta_0 (X)$. Note the magnified ($\times 20$) vertical axis for odd modes in right panel (b).}
%\label{fig:}
\end{figure*}

We now investigate the waveshifter in the spectral domain. To do so, the time response to a chirp signal between 10~kHz and 30~kHz is Fourier transformed to recover the spectral response. \textbf{Figure~2(a)} shows the real part of the out-of-plane velocity wavefield along the waveshifter, measured at two different frequencies, 16~kHz and 20~kHz. The experiment is compared to full-3D simulations at same frequencies for the same waveshifter (Figure.~2(b)), as well as for a uniform bent waveguide without the transformed medium (Figure.~2(c)). The $0^{th}$-order even mode of the waveguide efficiently converts into a new slanted ``$0^{th}$-order" mode of the anisotropic waveguide, with the wavefront oriented in the same direction as the incident mode. This is in stark contrast to the simulations performed in the bent waveguide without corrugation (Figure. 2(c)), where the incident mode converts into a combination of even and odd higher-order waveguide modes, and any information on the initial wavefront is lost.
%We refer to Figures. S2 \& S3 in [Supplementary Information] for simulations of dynamics of pulse propagation in the waveshifter for different bending angles $\theta$.

To better quantify the efficiency of the mode conversion in our waveguide shifter, we carry out a modal analysis of the measured and simulated fields inside the waveguide, before and after the bend \cite{shifter_water}. The modal decomposition is performed in the virtual space $(X,Y)$:
\begin{equation}
   \eta (X,Y) = \sum_{i=0}^{n}\eta_i(X) \psi_i(Y),
   %\label{eq:mode_decomp}
   \end{equation}
where $\eta_i (X)=\int_{-W/2}^{+W/2} \eta (X,Y) \psi_i(Y) dY$ refers to the integration across the width $W$ of the waveguide of the $i^{th}$-order transverse component, $\psi_i(Y)$. The first orders, even and odd, transverse modes are calculated in [Supplementary Information] for stress-free boundary conditions and shown in \textbf{Figure. 3(a)}. Before the bend, we simply have $(X=x,\ Y=y)$.  After the bend, the field $\eta (X,Y)$ is interpolated on a grid $(X, Y)$, using the inverse geometrical transformation $X=x$ and $Y = y - x \tan \theta $.

Figure.~3(b) and 3(c) show the measured and calculated first even and odd modes at f=16~kHz. Results confirm the efficiency of the energy transfer. The modal content of the incident signal, which is essentially the $0^{th}$-order even transverse mode, is perfectly preserved after deflection by the bend. This mode translates in the real space into the slanted mode seen in Figure.~2(c) with $k$-vector in the forward direction, $x$, but energy flow along the bent waveguide. Note that the scale for measured odd modes is 20 times smaller than that for even modes. After the bend, the conversion from incident mode to first odd mode due to the asymmetry of the designed structure, remains negligible. This contrasts with the empty waveguide (Figure.~3(d)) where the sharp change of direction couples the excitation to higher modes: the magnitude of first-order odd mode becomes comparable to that of mode 0 (Figure.~3(d)). This demonstrates the efficiency of our design, which deflects the $0^{th}$-order even mode into a slanted $0^{th}$-order mode, preserving the wavefront direction while preventing higher modes from being excited. The waveshifter is an interesting device which can be used as the building block of a variety of new functional components, including wave splitters, combiners  and invisibility cloak \cite{shifter_rahm}. This is illustrated in [Supplementary Information] with a cloaking device based on four waveshifters, arranged in a symmetric way around a diamond-shaped hole (see Figure. S5 in [supplementary Information]).

\section{Rotator}
\begin{figure*}
 \centering
\includegraphics[width = 16cm]{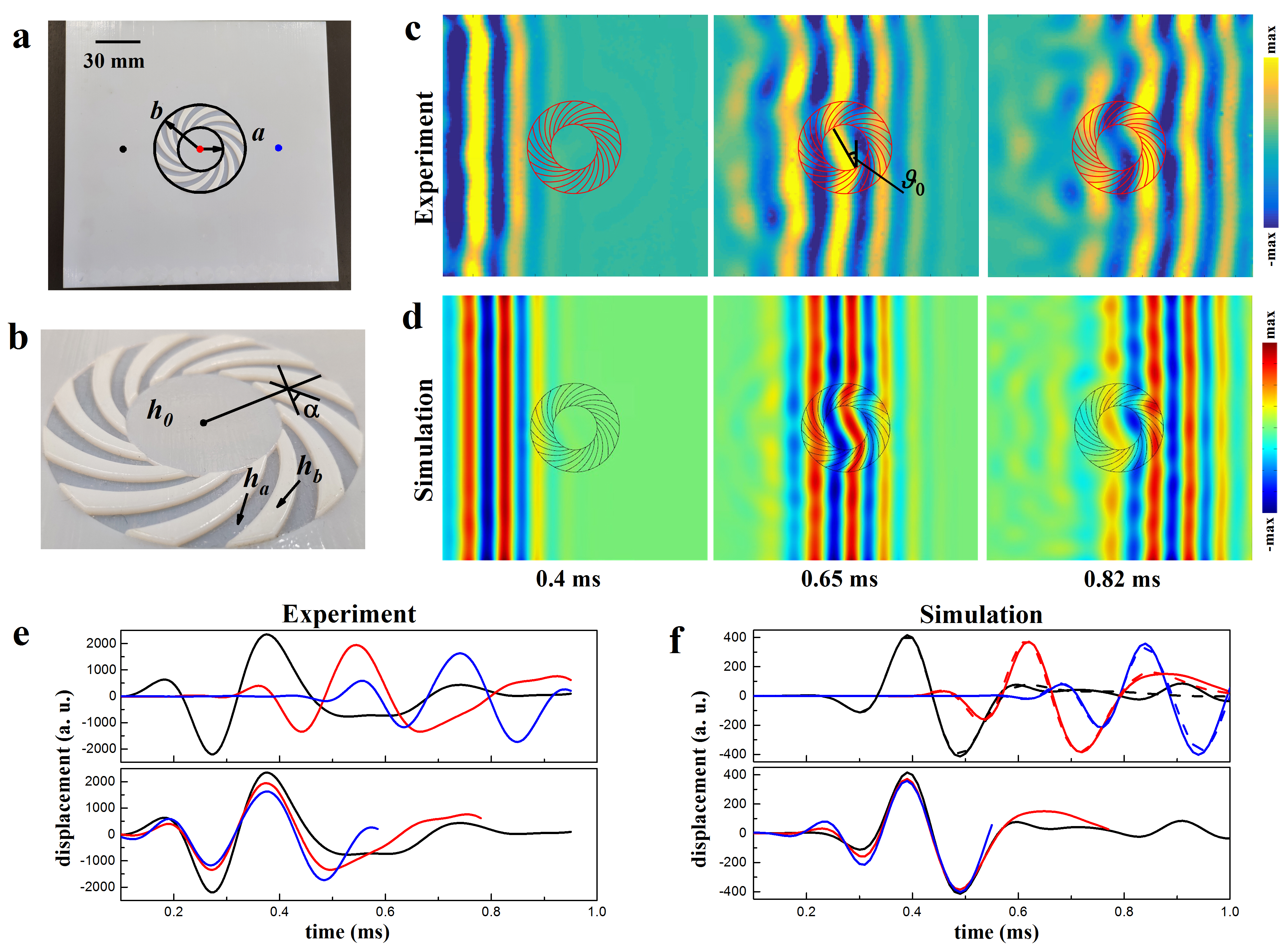}
  \caption{\textbf{Dynamics of pulse propagation in the rotator.} (a) Top view and (b) magnified view of the 3D-printed rotator of uniform plate thickness $h_0$=1~mm inside and outside the rotating annulus with inner radius $a$=15~mm and outer radius $b$=30~mm. The ring consists of spiraled corrugations of alternate heights $h_a$=2~mm and $h_b$=0.5~mm. (c) Experiment: Snapshots at time $t$= 0.4~ms (left), 0.65~ms (middle), and 0.82~ms (right) of the measured out-of-plane velocity field, which shows the plane wavefront of the Ricker pulse propagating before (left), during (middle) and after (right) the rotator (materialized by the red drawing). The rotation experienced by the elastic wave within the rotator is $\vartheta_0=30^{\circ}$. (d) Numerical simulations. (e) Experiment: temporal velocities at three spatial positions before, inside and after the rotator, marked by colored dots in (a). The red and blue curves in the lower panel are time-shifted to bring the three peaks in coincidence. (f) Same as (e) for numerical simulation: rotator plate (solid lines) and plain plate (dashed lines).}
 %\label{fig:experi}
\end{figure*}

% This cloaking effect is however restricted by construction to a single direction of incidence.
 By transposing the concept of the waveshifter from the Cartesian coordinate system to the polar coordinate system, we propose to design an isotropically %invisible
 functional device, the so-called wave rotator. By analogy with Equation.~(1), we define the following transformation
\begin{equation}
     \begin{cases}
      r = R \\
      \theta =  \vartheta_0 f(R) + \vartheta
       \end{cases},
 \end{equation}
where $f(R)$ is an arbitrary continuous function and $\vartheta_0$ the rotating angle of the device. Instead of the translation at fixed angle $\theta$ achieved by the waveshifter (Figure. 1a), this new coordinate transform rotates by an angle $\vartheta_0$ the wavefront incident from any direction, as it penetrates an annular region $a \leq r \leq b$. When the wave exits the annulus, the rotation effect is reversed and the wavefront direction is restored. A necessary condition on $f(R)$ is therefore $f(b) = 0$ and $f(a) = 1$. To push the analogy with the waveshifter a step further and allow a direct transposition of the design method proposed earlier, we assume that the Jacobian matrix associated with this new coordinate transform has the same form as in Equation.~(2), but defined this time in polar coordinate system:
\begin{equation}
  F =\left(
 {\begin{array}{ccccc}
    \scriptstyle \frac{\partial r}{\partial R} & \scriptstyle \frac{1}{R}
    \frac{\partial r}{\partial \vartheta}\\
  \scriptstyle -R \frac{\partial \theta}{\partial R} & \scriptstyle \frac{\partial \theta}{\partial \vartheta}\\
  \end{array}}
       \right)
  =\left(
 {\begin{array}{ccccc}
   1 & 0\\
   t ^\prime & 1\\
  \end{array}}
       \right),
\end{equation}
where $t^\prime$ is a constant to be defined. A necessary condition to satisfy this equality is
\begin{equation}
-R \frac{\partial \theta}{\partial R}=t^{\prime}.
\end{equation}
From this condition, we obtain
\begin{equation}
\left\{
\begin{array}{lr}
f(R) =  \frac{\ln(b/R)}{\ln(b/a)}\\
t^{\prime} = \frac{\vartheta_0}{\ln(b/a)}
\end{array}
\right .
\end{equation}

For given rotating angle $\vartheta_0$ and radii $a$ and $b$, the Jacobian matrix $F$ is constant, independent of space coordinates and its determinant is unity so that the transformation preserves the volumes. Under these conditions, the change of material parameters which realizes the distortion of the wave of Equation.~(7), is obtained by following step by step the procedure proposed to design the waveshifter. The rigidity tensor $\overline{D}$ is again given by Equation.~(3), while the mass density remains unchanged $\rho=\rho_0$. We note once more that $\rho$ is a factor of second time derivative of the Kirchhoff-Love equation, which therefore ensures the form-invariance of the kirchhoff-Love equation in the time domain. The anisotropy is introduced in the same way, by alternating layers with different flexural rigidities. The angle $\alpha$ formed by these layers with the local polar frame is constant, as it was with the tilted Cartesian frame of the waveshifter. By definition, the curve defined in polar coordinate by a constant tangential angle $\alpha$ is a logarithmic spiral, $r=a.e^{k(\theta-\beta)}$, with $k= -\tan \alpha$, and $\beta$
is an arbitrary initial angle for $r=a$. The angle $\alpha$ is given by Equation.~(5), where $t$ must be replaced by $t^{\prime}=\frac{\vartheta_0}{\ln(b/a)}$. The subwavelength anisotropic structure is realized by choosing $N=24$ initial angles
\begin{equation}
\beta =0, \frac{2\pi}{N},...(N-1)\frac{2\pi}{N},
\end{equation}
which defines $N$ logarithmic spirals (see \textbf{Figure 4(a) \& 4(b)}). Regions between 2 successive spirals define regions with alternating flexural rigidities $D_a$ and $D_b$, as defined in Equation.~(5). Following the method used for the design of the waveshifter, this is implemented practically by varying the thicknesses $h_a$ and $h_b$ of the plate in these regions. The resulting structure is similar to \cite{cloaking_darabi,fisheye,lenses_climente,Grin_lense}.

A rotator with rotation angle $\vartheta_0=30^{\circ}$ has been designed and 3D-printed on a stiff photo-resin in a 18~cm~$\times$~18~cm square plate with thickness $h_0$=1~mm. The inner and outer radius for the rotating annulus are $a$= 15~mm and $b$= 30~mm, respectively. A close-up on the rotator (Figure. 4(b)) shows the spiral-like corrugation with alternating thicknesses $h_a$=2~mm and $h_b$=0.5~mm. We propagate a short pulsed plane wave (Ricker pulse centered at 4kHz) across the rotating device. Figure 4(c) presents successive snapshots of the wavefield vertical velocity at three different times, while a complete movie is available in [Supplementary Information]. This shows clearly how the incident wave acquires a $30^ {\circ}$ anticlockwise twist as it penetrates the rotator. As it exits the device, the wavefront is rapidly restored after about one wavelength and the plane wave continues its journey as if nothing had happened. Note that this illusion that nothing has happened, which conveys the concept of a cloaked event as proposed with a space-time cloak in electromagnetism ten years ago \cite{McCall_2010},
%of \textcolor{red}{nothing happened}
%total invisibility
is also achieved in the backward direction where almost no perceptible energy is being scattered. Besides, rotational symmetry ensures that this illusion %invisibility
is achieved from any direction the device is looked at. Full-3D time-domain simulations confirms this behavior (Figure. 4(d)).
Actually, not only the wavefront is preserved but the temporal shape of the pulse does not experience any spatial or temporal distortion. This is demonstrated in Figure. 4(e) and 4(f) where we compare the temporal pulse profile measured at three different positions, on both sides and inside the rotator, as marked by the colored dots in Figure. 4(a). By time-shifting the peaks, we show good temporal overlap for the experiment and excellent coincidence for the numerical simulations: the pulse profile has been preserved. Even more striking, we found that no phase delay is accumulated during propagation across the rotating device. This is shown by comparing the pulse after crossing the rotator (full red line) to a pulse measured at the same position in a plain plate without the rotator device (dashed red line). The perfect overlap of the two time signals demonstrates that rotating effect
%cloaking
provided by the rotator is perfect. This is in stark contrast with other cloaks based on resonant dispersive structures, where pulse experiments would inevitably disclose the presence of the device \cite{optic_delay}.

\section{Mirage effect}
Actually, the wave rotator is not a cloaking device per se as it does not hide an object. It presents however the surprising ability to create a mirage effect, by giving the illusion that the object inside the device is located in a deceptive position. This is tested  numerically here for flexural waves in the time domain, based on the proposed design. We clamp, within the annulus of the wave rotator studied earlier, a rectangular obstacle at an angle $\vartheta_0=30^\circ$ with respect to the wavefront of the incident pulsed plane wave. The field distribution of the flexural mode is recorded at different times, as shown in \textbf{Figure. 5(a)}. The tilted obstacle preferentially reflects back the elastic field at an angle $-2\vartheta_0$. When the rotator structure is added around the obstacle, the field is now backscattered horizontally, as if the rectangular object was perfectly aligned with the incident wavefront (Figure. 5(b)). This is confirmed by comparing the field reflected by the tilted obstacle in the presence of the cloak (Figure. 5(b)) and the field reflected by the obstacle in a vertical position without the cloak (Figure. 5(c)), which shows the same spatial distribution. A more quantitative analysis is proposed in Figure.~5(d) by comparing the scattering diagrams at time step 1.3~ms and at a distance of 60~mm from the center of the rectangular scatterer. Good overlap is found in the backward direction ($0^\circ$) of the scattered-field distributions induced by the bare vertical scatterer (blue line) and the cloaked tilted scatterer (red line), while in contrast, the bare tilted scatterer reflects the wave at $-2\vartheta_0=-60^\circ$ (black line). One can say that the object has deceitfully been straightened in vertical position by the rotator, with the impression for a distant observer that scattered wave comes from an unexpected direction.

\begin{figure*}
 \centering
\includegraphics[width = 16cm]{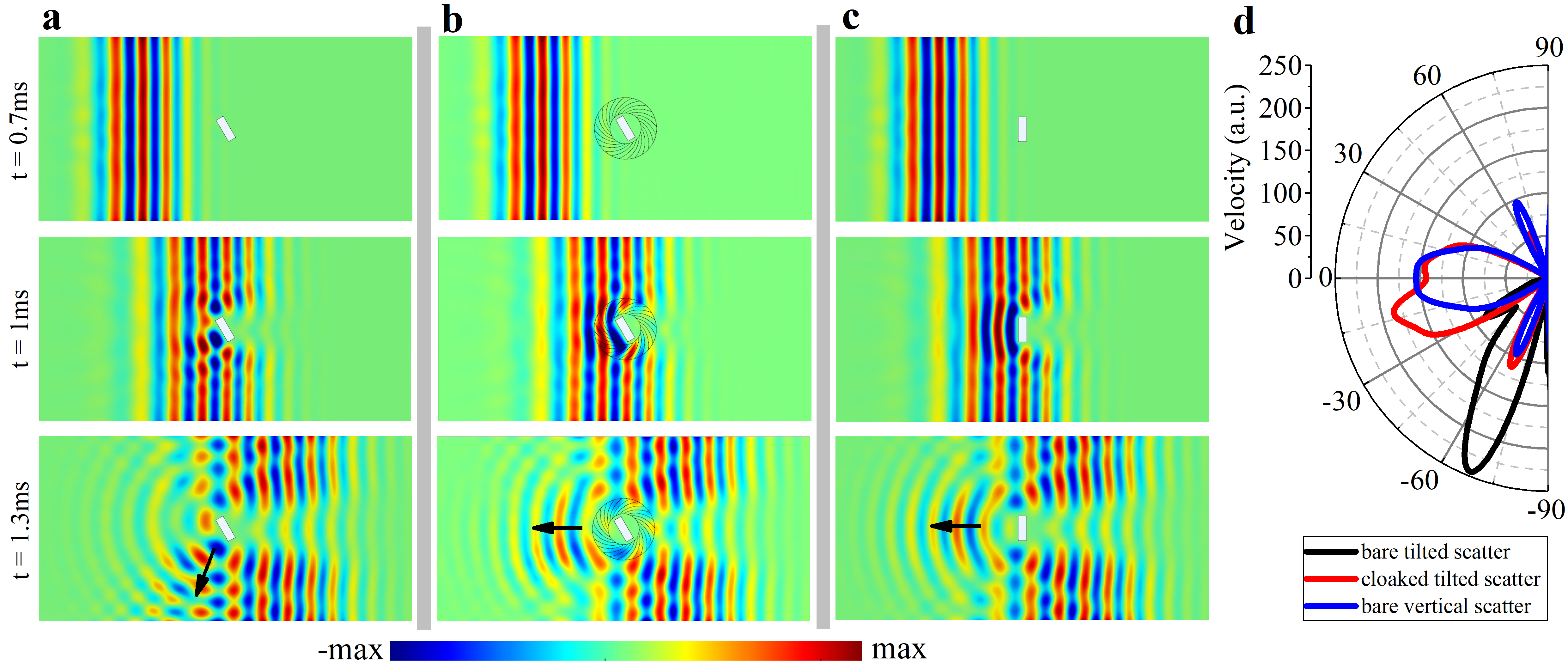}
\caption{\textbf{Dynamic mirage effect in transient regime}. Full 3D elastic-wave simulations showing the propagation across the wave rotator of a flexural plane wave (Ricker pulse centered at 4 kHz) launched from left edge. (a) Homogeneous plate with a clamped rectangular scatter tilted at $30^\circ$ with respect to the wavefront of the incident plane wave; (b) Plate with the transformed rotator enclosing the same tilted scatter; (c) Plate with a scatter in vertical position with no rotator. (d) %\textbf{Scattering diagram of the mirage effect}.
Scattered-field angular distribution calculated at time step 1.3ms and at a distance of 60~mm from the center of the scatterer for (blue line) the bare vertical scatterer; (red line) the cloaked  scatterer tilted at $30^\circ$ with respect to the wavefront of the incident plane wave; and (black line) the bare tilted scatterer. $0^\circ$ represents the backward direction.
}
%  \label{fig:experi}
\end{figure*}

\section{Conclusion}
In conclusion, we have shown that coordinate transformation can be adapted to design new devices for flexural waves. This leads us to an interesting strategy for the design of transformed elastic devices. We proposed and successfully tested experimentally a waveshifter and a wave rotator. To shape the anisotropic transformed space, we have taken the option to carefully engineer a single material by corrugating the surface of the plate on a subwavelength scale, a design well-suited for practical realization. The magic of the coordinate transformation manifests itself in the simply-designed reflectionless waveshifter, where the $0^{th}$-order elastic mode maintains the direction of its wavefront beyond the bend, while its energy flow is deviated in a different direction. We demonstrate that both devices work equally well for short pulses, with virtually no spatial or temporal dispersion. We note that this can be attributed to that the geometric transformation is area-preserving. The rotator turns out to be a truly transparent device since the phase delay accumulated across the device is the same as it would be in free space. To the best of our knowledge, this has never been observed, with supposedly invisible devices. The usefulness of analogies with flexural waves in plates, but also with wave optics and water waves for the control of surface seismic waves was pointed out in \cite{seismic_wave}. We believe that our design approach can offer an interesting route to the control of e.g. surface Rayleigh waves in soils, structured in a similar fashion to elastic plates, using an alternation of trenches and walls, in order to control quakes in the time domain.

% Experimental section

\section{Experimental Section}
\emph{Measurement setup}: A 3D-printer based on nanoparticle jetting technology (Xjet, Carmel 1400) has been used to fabricate the waveshifter waveguide of Figure.~1(b). The device was printed on zirconia ($Z_rO_2$), a ceramic with  the following elastic parameters: Young's modulus $E$=207~GPa, mass density $\rho$=6040~kg/m\textsuperscript{3}  and Poisson's ratio $\nu$ =~0.32.
%In the frequency range of interest (15 to 25 kHz), the plate thickness is much smaller than the wavelength (15mm at 25 kHz), whereas the width of the waveguide (10 mm) is comparable to half a wavelength, only the fundamental mode (mode 0) exists, which possesses a mode profile close to a plane wave.
A piezoelectric diaphragm (Murata 7BB-12-9) located on the flat arm of the waveguide is used to excite the fundamental mode of the waveguide. Both ends of the waveguide are covered with blu-tack on both sides to reduce reflections.
%on the measured grid data for a better visualization of the wavefield on the plate.

The rotator was manufactured using PolyJet (Stratasys Objet 24), a technology based on photo-polymer 3D printing. Here we used Vero PureWhite™ RGD837, a stiff resin with Young's Modulus E= 2.5GPa, mass density $\rho$= 1180 kg/m\textsuperscript{3} and Poisson's ratio $\nu$ = 0.25.
%In the frequency range of interest (4 kHz), the wavelength (26mm) is much larger than the plate thicknesses and widths of the alternating layers, which ensures the accuracy of the thin plate approximation and homogenization of metamaterial structures.
Eighteen piezoelectric diaphragms (Murata 7BB-12-9) were bonded along one edge of the 180~cm $\times$ 180~cm square plate. All transducers are excited simultaneously with the same signal to generate a plane wave. Blu-tack was also used to reduce reflection at the edge on the plate.
In both cases, a Ricker pulse was generated at each transducer by an arbitrary function generator (Agilent 33220A) with the addition of a high-voltage amplifier. A laser vibrometer (Polytec sensor head OFV534, controller OFV2500) was scanned on the flat surface of the device to measure the spatio-temporal velocity field of the flexural waves (1mm step grid for the waveshifter and 2~mm step grid for the wave rotator).
The images are processed using a Hampel filter and a cubic interpolation.

\emph{Ricker pulse}: We recall that a Ricker pulse is the second derivative of a Gaussian function and is defined  in the time domain by:
\begin{equation}
A= \left[1-2\pi^2{f_0}^2(t-1/f_0)^2\right]e^{-\pi^2{f_0}^2(t-1/f_0)^2}.
\label{eq:ricker}
\end{equation}
Note that this zero-mean symmetric pulse is solely defined by a single parameter, its most energetic frequency $f_0$.

\emph{Numerical simulations}: All 3D full-wave simulations were conducted with the Solid Mechanic Module of the finite element software COMSOL Multiphysics 5.3. Low-reflection boundary were imposed in the frequency domain, on the left and right ends of the waveshifter and rotator plate, and on the outer boundaries of the cloaking plate. The largest mesh-element was set to be smaller than one-tenth of the lowest wavelength. A finer mesh was used where the domain geometry changes abruptly.

\medskip
\textbf{Supporting Information} \par %Please delete the Suppporting Information statement if it is not applicable. Please supply Supporting Information in another file. Supporting information should not be provided in .tex format
Supporting Information is available from the Wiley Online Library or from the author.

% Acknowledgements
\medskip
\textbf{Acknowledgements} \par %delete if not applicable))
The authors thank Agnes Maurel from the Langevin Institute, ESPCI-CNRS, for fruitful discussions on homogenization of elastic systems, Avi Cohen from X-jet for the waveshifter fabrication, and Shai Ingber from SU-PAD for the rotator fabrication. This research was supported in part by The Israel Science Foundation (Grants No. 1871/15, 2074/15 and 2630/20) and the United States-Israel Binational Science Foundation NSF/BSF (Grant No. 2015694). P. S. is thankful to the CNRS support under grant PICS-ALAMO.

\medskip

\bibliographystyle{MSP}
\bibliography{mybibligraphy}



\section*{Supplementary material}

# Supplementary Information: Pulse dynamics of flexural waves in transformed plates

Kun Tang[1], Chenni Xu[1,2], Sébastien Guenneau[3], & Patrick Sebbah[1]

[1] Department of Physics, The Jack and Pearl Resnick Institute for
Advanced Technology, Bar-Ilan University, Ramat-Gan 5290002, Israel
[2] Zhejiang Province Key Laboratory of Quantum Technology and Device,
Department of Physics, Zhejiang University, Hangzhou, 310027, Zhejiang, China and
[3] UMI 2004 Abraham de Moivre-CNRS, Imperial College London, London SW7 2AZ, United Kingdom

## PERFORMANCES OF THE DESIGNED WAVESHIFTER WITH DIFFERENT BENDING ANGLES $\theta$

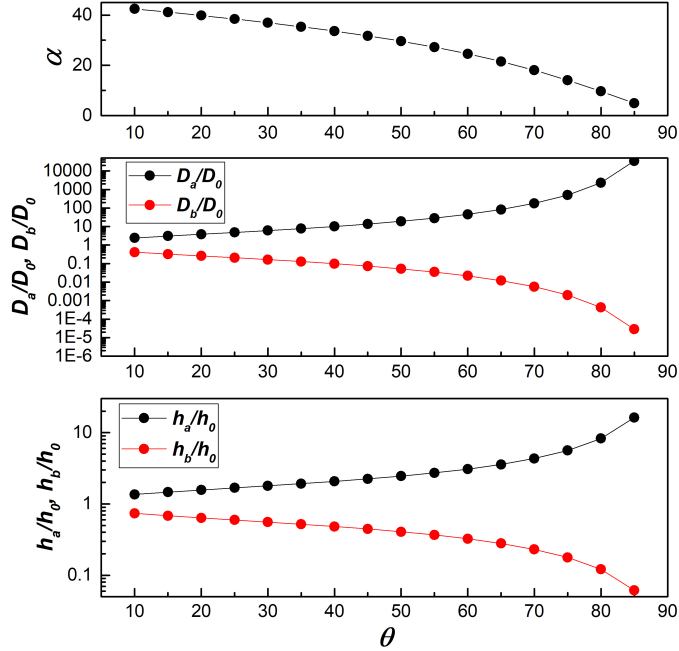

FIG. S1: **Geometric parameters of the waveshifter dependence on bending angle $\theta$.** Plot of $\alpha$, $D_a$ ($D_b$), $h_a$ ($h_b$) vs. $\theta$, using Equation. (5) of the main article.

In the experimental investigation and numerical analysis presented in the main text, the bending angle $\theta$ of the waveshifter was chosen to be $20°$. Here, we check the validity of our approach for larger values of $\theta$. The dependence on $\theta$ of the plate rigidities $D_a$, $D_b$, the rotating angle $\alpha$ and the plates thicknesses $h_a$, $h_b$, are calculated following Equation. (5) and shown in Figure. S1, see also Figure. 1(b). When $\theta$ varies between 0 and $90°$, $\alpha$ decreases gradually, while $D_a$ and $h_a$ diverge rapidly near $90°$ (notice the logarithmic scale in Figure. S1). This indicates that much stronger anisotropy is required for sharper bending angle $\theta$.

Following the method presented in the main text, we design a waveguide with the same geometry as the waveshifter shown in Figure. 1, but with different bending angles $\theta$, ranging from $30°$ to $70°$. The corresponding values of $\alpha$, $h_a$ and $h_b$ are directly obtained from Equation. (5). The out-of-plane velocity field distribution is calculated at $16 kHz$ and shown in Figure. S2 with a comparison to the uniform bent waveguide without corrugation. For values of $\theta$ between $30°$ and $60°$, the $0^{th}$-order mode of the waveguide is efficiently transmitted beyond the bend, with the direction of its wavefront well preserved. In contrast, the same incident mode is seen to convert into a combination of even and odd higher-order modes in the uniform waveguide without corrugation. Beyond these values ($\theta \geq 70°$), the waveshifter looses its characteristics: the wavefront direction is no longer preserved leading to mode conversion and

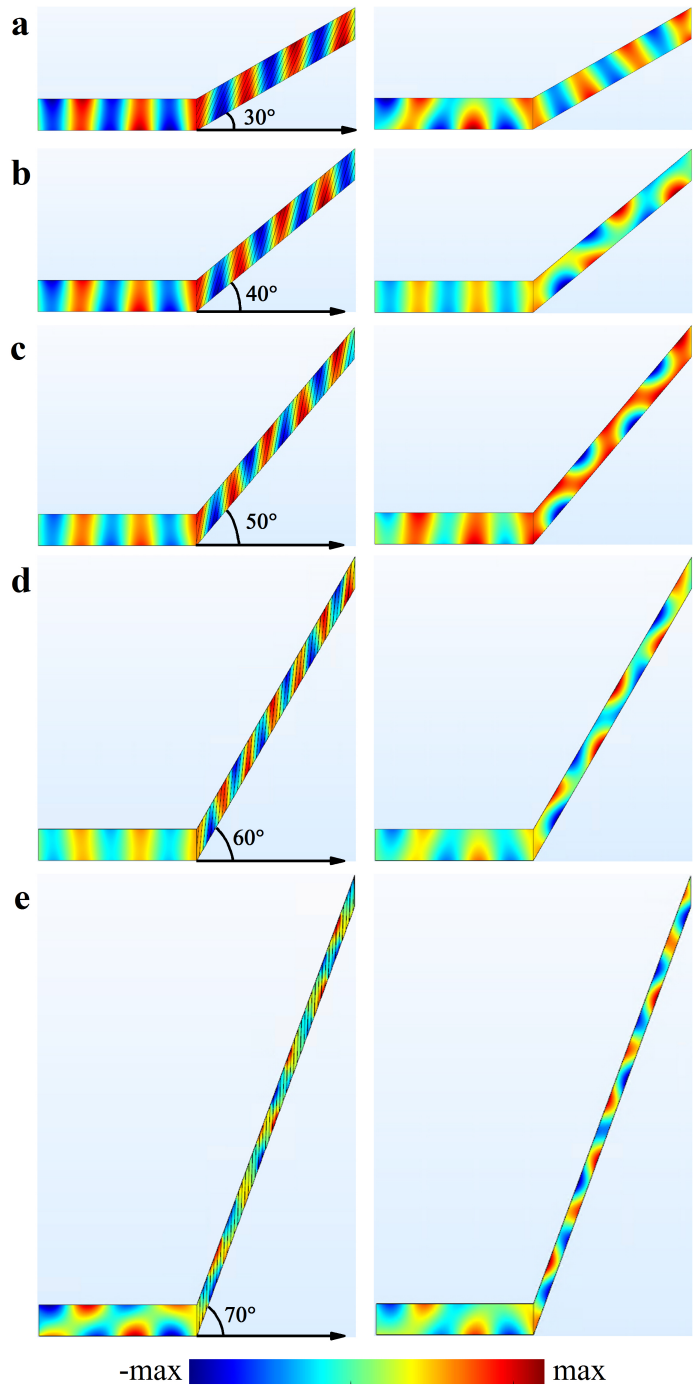

FIG. S2: **Waveshifter with different bending angles $\theta$.** Left column: Simulated out-of-plane velocity field at $f = 16$ kHz for the waveshifter with different bending angles (a) $\theta = 30°$, (b) $\theta = 40°$, (c) $\theta = 50°$, (d) $\theta = 60°$ and (e) $\theta = 70°$. Right column: Numerical simulations for the corresponding empty waveguides without any corrugation.

back-reflection. At these angles, the homogenization formula breaks down, which assumes a low, or at least moderate, contrast in material parameters.

We present in Figure. S3 snapshots of the velocity field distribution at three different times in response to a short

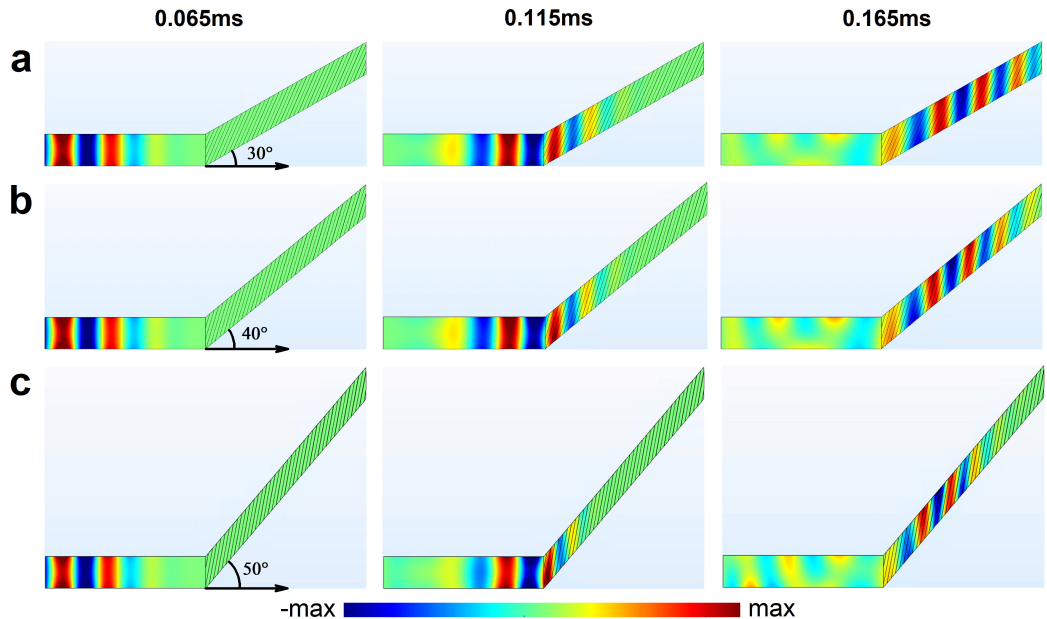

FIG. S3: **Dynamics of pulse propagation in the waveshifter with different bending angles $\theta$.** (a) Full-3D numerical simulations: Snapshots of the out-of-plane velocity field calculated in response to a Ricker pulse with central frequency 20 kHz, at times 0.065 ms (left panel), 0.115 ms (middle panel), and 0.165 ms (right panel), for waveshifters with bending angles (a) $\theta = 30°$, (b) $\theta = 40°$, and (c) $\theta = 50°$.

incident pulse, for bending angles ranging from $30°$ to $50°$. The pulse is seen to propagate smoothly across the waveguide bend with negligible back-reflections, while the wavefront remains vertical, in the same direction as the incident wave. This confirms the performance for short pulses of our designed waveshifters for large bending angles, up to $50°$. For larger angles $\theta$ and larger contrast in material parameters, local resonances occur in the long wavelength limit and the resulting effective parameters become dispersive. The metamaterial waveguide no longer approximates a dispersionless (frequency independent) transformed medium, and this has a deleterious impact on the metamaterial efficiency. In fact, the larger the contrast in material parameters, the more dispersion in effective parameters, and thus the more impact on the pulse propagation. The choice of $\theta = 20°$ in the experimental demonstration presented in the main text is a good compromise between experimental constraints and large values of $\theta$, the main restriction being the resolution of the 3D printing machine and the thickness $h_b$, which decreases rapidly with $\theta$ (see Figure S1).

## BROADBAND PERFORMANCE OF THE DESIGNED WAVESHIFTERS

In order to estimate the working bandwidth for designed waveshifters, we calculate the contribution of mode 0 to the total field for the waveshifter with bending angle of $\theta = 20°$, in transmission $X > 0$. To do that, we define $\mid \eta_0 \mid = \int dX \mid \eta(X) \mid$, $\mid \eta \mid = \int dX \int_0^H dY \mid \eta(X, Y) \mid$, where $H$ is the width of the waveguide, $\mid \eta(X, Y) \mid$ is interpolated on a grid $(X, Y)$ using the transformation $X = x$ and $Y = y - x \tan\theta$. The integrals are performed for $X > 0$ and we denote $\mid \eta_0 \mid / \mid \eta \mid$ the weight of mode 0 in the transmission. As shown in Figure .S4, we compare the contribution of mode 0 in transmission for waveshifter with corresponding empty bent waveguide, under Perfectly Matched Layer (PML) boundary condition at the entrance and exit of the waveguide. The mode 0 remains largely dominant (above 90%) for all frequencies ranging from 0.1kHz to 30kHz (i.e. with a fractional bandwidth of 200%, black solid curve), while it decreases linearly for the corresponding empty bent waveguide (red solid curve).

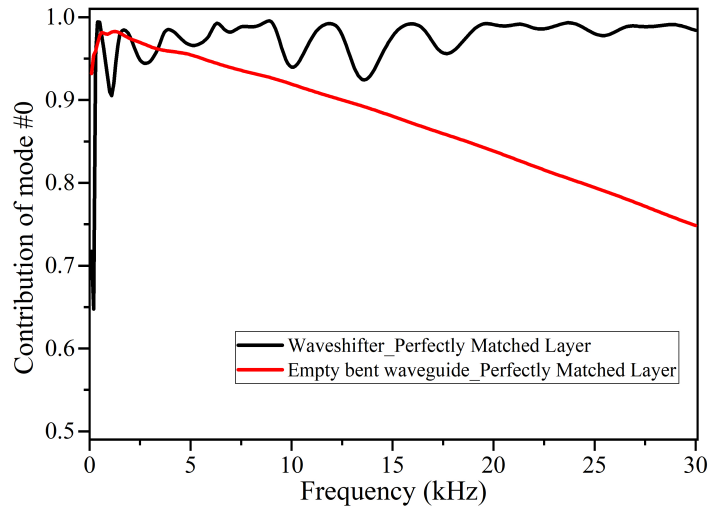

FIG. S4: Contribution of mode 0 to the total field in transmission $X > 0$ as a function of frequency for waveshifter (black color) and corresponding empty bent waveguide (red color) with bending angle of $\theta = 20°$, supplied with Perfectly Matched Layer in the simulations.

## CLOAKING

The waveshifter is an interesting device which can be used as the building block of a variety of new functional components, including wave splitters, combiners and invisibility cloak [1]. This is illustrated here with a cloaking device based on four waveshifters, arranged in a symmetric way around a diamond-shaped hole (see Figure. S5). Such a cloaking device has been proposed in [2] for continuous light waves and is demonstrated here for pulsed elastic waves. We use the parameters calculated earlier, $\theta$, $h_a$, $h_b$, $h_0$, and $\alpha$, but instead of limiting the transformed region to a waveguide geometry, we simulate an incident pulse (Ricker pulse centered at 20 kHz) with a 160 mm-wide Gaussian wavefront propagating in a wide corrugated area around the diamond-shaped hole. Figure S5 compares the field distribution of the vertical elastic velocity resulting from the scattering by a bare hole with stress-free boundaries without and with the cloaking device, at three different time steps. The Gaussian wavefront splits as it hits the hole and the wavefront rapidly breaks apart. With the cloak however, the wavefront is maintained while following the edges of the obstacle and recombines after the hole. Beyond the hole, the initial Gaussian profile is restored. In addition to the invisibility effect, the cloak itself is invisible, with negligible back reflection. We checked that the temporal elongation of the pulse is solely due to flexural waves dispersion, as it would occur naturally in a plain plate without the obstacle.

Following this design, we construct a cloaking device for different values of $\theta$. We use the parameters calculated in Figure. S2, $h_a$, $h_b$, $h_0$, and $\alpha$, for different values of $\theta$. A Ricker pulse centered at 20 kHz with a 160 mm-wide Gaussian transverse profile is propagated in the direction of the hole. The field distribution of the out-of-plane velocity computed at time step 0.38ms, is shown in Figure. S6 for $\theta = 20°$, $\theta = 30°$, $\theta = 40°$ and $\theta = 50°$, for the diamond-hole with and without the cloaking corrugation. We find that the outgoing wave points in the same direction as the incident wave and that the incident wavefront is well restored for metamaterial cloaks with bending angles up to $\theta = 40°$. This is in stark contrast with the field distributions resulting from the scattering by a bare hole ($2^{nd}$ row in Figure. S6), where the incident plane wavefront is seen to scatter off into multiple directions, and any information on the initial wavefront is lost. The cloaking device is therefore clearly effective to cloak even large diamond-shape holes, at least for diamond angles up to $\theta = 40°$. For $\theta \geq 50°$, the metamaterial cloak still refocuses efficiently the wavefront in the incident direction, but starts to yield significant wavefront distortions.

To better quantify the cloaking efficiency, we perform a spatial Fourier transformation to obtain the velocity amplitude in wave-vector ($k$) space ($3^{rd}$ and $4^{th}$ row in Figure. S6). The simulated wave field in $k$ space for metamaterial cloak is mainly localized around two sparkling lines, similar to wave propagation inside the plain plate, which indicates most of the energy points at the forward direction. However, the scattered wave by the bare hole is composed of a wider range of wave vectors located within an annulus regime. It means that the scattered wave no longer remains a forward plane wavefront and orient in multiple directions. This confirms that the metamaterial cloak refocuses

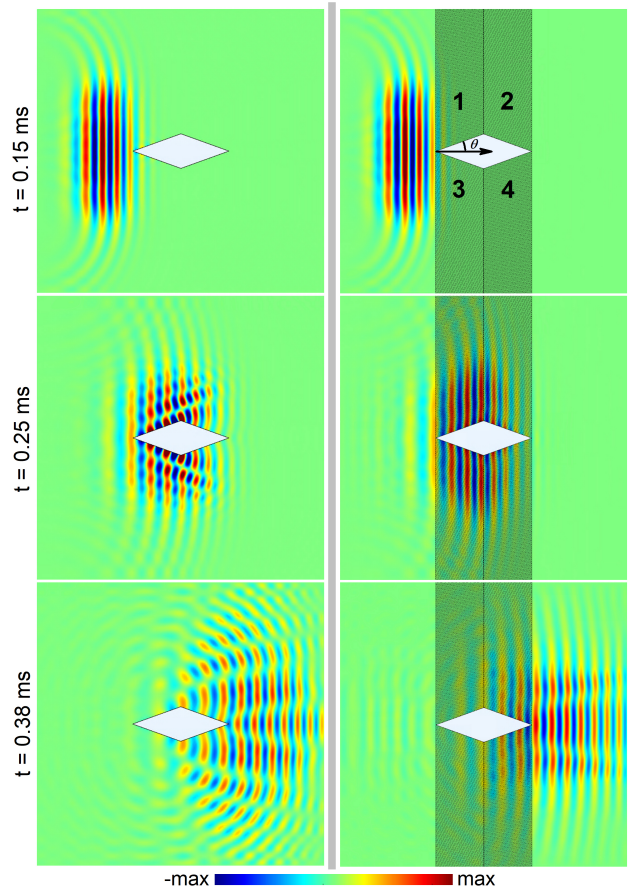

FIG. S5: **Cloaking flexural waves with 4 waveshifters**. Full-3D elastic-wave transient simulations of elastic field distribution resulting from the propagation of a pulsed gaussian wave (Ricker pulse centered at 20 kHz) with 160 mm-wide transverse profile, at times t=0.15 ms, t=0.25 ms, t=0.38 ms. (left panel) Bare diamond-shaped hole; (right panel) diamond cloak. The cloak is composed of 4 corrugated regions around a diamond-shaped hole with θ=20°, as defined in the figure. Geometric parameters of the corrugated region 1 are identical to those of the waveshifter of Figure. 1. Region 3 is the mirror image of Region 1 with respect with the horizontal axis. Regions 2 and 4 are mirror images of Region 1 and 3 with respect to the vertical axis. One notes that scattering off the hole is nearly suppressed by the corrugated regions at time step $t = 0.38$ms: The cloak flattens the concentric wavefronts emanating from the hole. See Figure. S4 in [Supplementary Information] for same simulations performed for holes and cloaked holes with larger $\theta$.

efficiently the wavefront in the incident direction for bending angles up to $\theta = 50°$.

The waveshifter can thus be viewed as a building block for certain types of diamond shaped cloaks, such as shown in Fig. S6. Nonetheless, in the similar way to the waveshifter, a diamond cloak design is constrained by the requirement of a moderate contrast in material parameters that approximate the transformed plate medium, which is dictated by the angle $\theta$ inside the diamond shaped hole. The increasing dispersion of the effective medium induced by the increasing contrast in material parameters (here the layers' thicknesses), leads to an increased distortion of the Ricker pulse for large $\theta$, as becomes clear for $\theta = 50°$. In fact, the physical phenomenon of effective dispersion induced by large contrast in materials parameters of periodic structures occurs for any type of waves, so similar constraints exist for acoustic and electromagnetic cloaks: The price to pay for a cloak suppressing the scattering of a large object is the requirement of a large anisotropy of the transformed medium, which will then be approximated by periodic structures with large contrast in material parameters. Therefore such cloaks might work well in theory throughout a large frequency band when using homogenization approaches to mimic the transformed media, but their efficiency will in practice remain moderately low for Ricker pulses due to the inherent dispersion in the effective parameters.

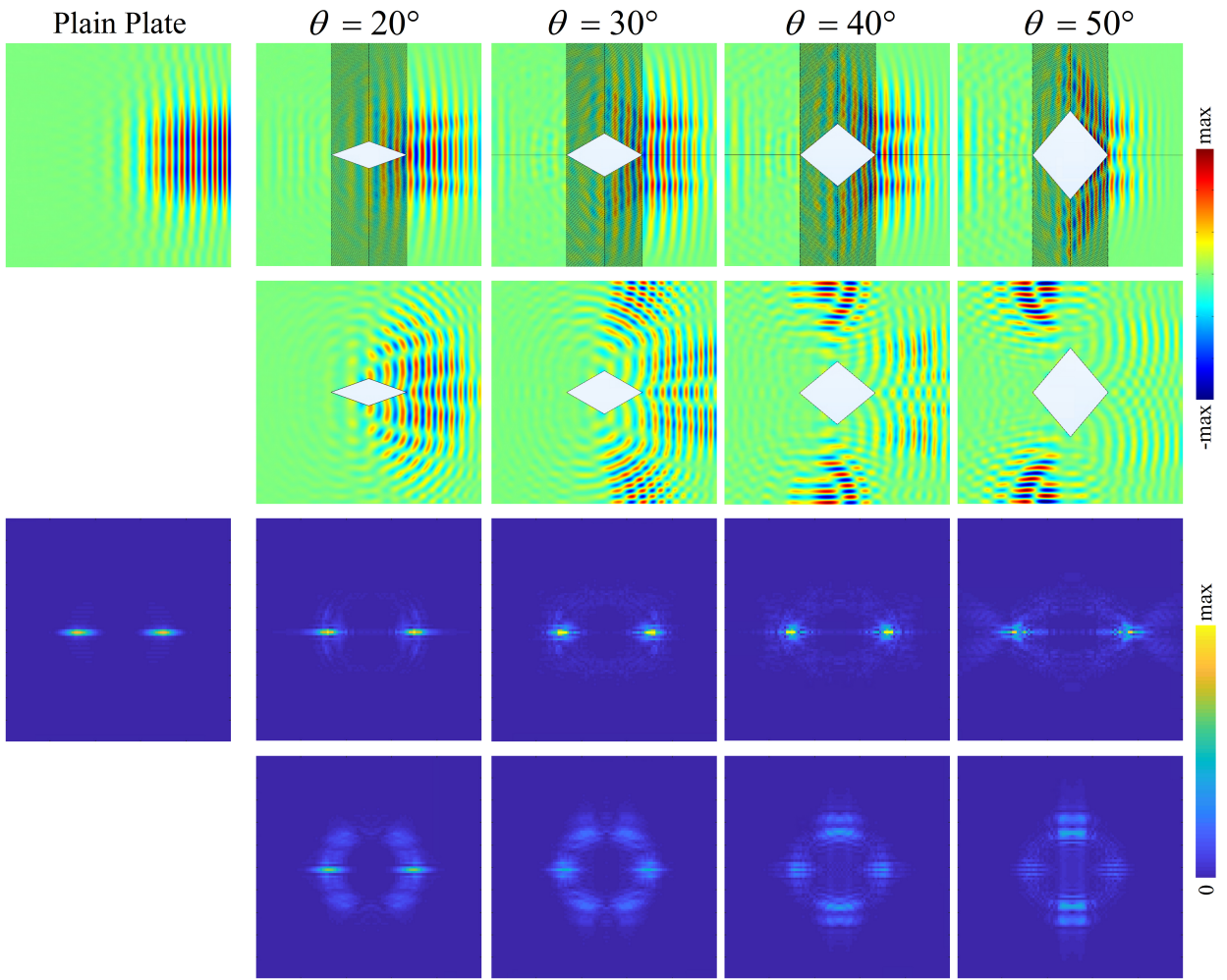

FIG. S6: **Diamond-shape cloak with increasing angle $\theta$.** Full-3D elastic-wave transient simulations of elastic field distribution at time step 0.38 ms: ($1^{st}$ column) for the plain plate; ($1^{st}$ row) for the diamond cloak composed of four corrugated regions, ($2^{nd}$ row) for the bare diamond-shaped hole, with bending angles ($2^{nd}$ column) $\theta = 20°$, ($3^{rd}$ column) $\theta = 30°$, ($4^{th}$ column) $\theta = 40°$, ($5^{th}$ column) $\theta = 50°$. Spatial Fourier transformation is performed on the velocity field shown in the $1^{st}$ and $2^{nd}$ rows to obtain the velocity amplitude in wave-vector space shown in $3^{rd}$ and $4^{th}$ rows.

## EIGENMODES OF THE WAVEGUIDE

Here we provide a short explanation on calculating the flexural eigenmodes supported by homogeneous plate with stress-free boundaries. The details can be found in ref. [3]. As Kirchhoff-Love plate equation is of fourth order, there exist two sets of modes at each frequency

$$w^{(e)}(y) = A\left[ \cosh\left(\frac{\chi_m W}{2}\right)\cosh(\chi_p y) - \frac{k^2\nu - \chi_p^2}{k^2\nu - \chi_m^2}\cosh\left(\frac{\chi_p W}{2}\right)\cosh(\chi_m y) \right] \tag{1}$$

$$w^{(o)}(y) = A\left[ \sinh\left(\frac{\chi_m W}{2}\right)\sinh(\chi_p y) - \frac{k^2\nu - \chi_p^2}{k^2\nu - \chi_m^2}\sinh\left(\frac{\chi_p W}{2}\right)\sinh(\chi_m y) \right] \tag{2}$$

Where $\chi_p = \sqrt{k^2 + K^2}$, $\chi_m = \sqrt{k^2 - K^2}$, $K^2 = \omega\sqrt{\rho h/D}$, and $A$ is a normalization constant such that $\int_{-W/2}^{W/2} dy \mid w^{(o,e)}(y) \mid^2 = 1$. The dispersion relations $\omega(k)$ can be derived by solving the following transcendental equations

$$\left[K^2 + (1-\nu)k^2\right]^2 \chi_m \tanh\left(\chi_m W/2\right) = \left[K^2 - (1-\nu)k^2\right]^2 \chi_p \tanh\left(\chi_p W/2\right) \tag{3}$$

for even modes, and

$$\left[K^2 + (1-\nu)k^2\right]^2 \chi_m \coth\left(\chi_m W/2\right) = \left[K^2 - (1-\nu)k^2\right]^2 \chi_p \coth\left(\chi_p W/2\right) \tag{4}$$

for odd modes.

It yields the wavenumber $k_i^{(o,e)}$ associated to each mode $w^{(o,e)}$ at frequency $\omega$. The $y$-dependence of the first two even and odd eigenmodes is shown in Fig. 3(a). The mode profile of first even mode (mode 0) is close to a plane wavefront.

### ANIMATIONS OF THE PULSES PROPAGATION INSIDE WAVESHIFTER AND ROTATOR.

The pulse dynamic propagation of the measured and simulated wave patterns inside waveshifter and rotator are shown in the enclosed videos.

- Video: "waveshifter-experiment", "waveshifter-simulation".
  Measured and simulated time evolution of the spatial distribution of the displacement for wave propagation through designed waveshifter. The incident pulse is a Ricker pulse with peak frequency $f_0 = 20kHz$.

- Video: "rotator-experiment", "rotator-simulation".
  Measured and simulated time evolution of the spatial distribution of the displacement for wave propagation through designed rotator. The incident pulse is a Ricker pulse with peak frequency $f_0 = 4kHz$.

- Video:"mirage effect-vertical scatter","mirage effect-rotated scatter", "mirage effect-rotated scatter with outside rotator".
  Simulated time evolution of the spatial distribution of the displacement through bare vertical scatter, rotated scatter, and rotated scatter with outside rotator. The incident pulse is a Ricker pulse with peak frequency $f_0 = 4kHz$.

- Video:"1D cloak-void","1D cloak-void with a cloak".
  Simulated time evolution of the spatial distribution of the displacement for wave propagation through a bare diamond-shaped void or void coated with a one-dimensional cloak. The incident pulse is a Ricker pulse with peak frequency $f_0 = 20kHz$.

———————



\end{document}